\documentclass[acmsmall,screen,authorversion,nonacm]{acmart}
\usepackage[T1]{fontenc}
\usepackage{tcolorbox}
\usepackage{pythonhighlight}
\usepackage{amsmath}
\usepackage{graphicx}
\usepackage{xcolor}
\usepackage{enumitem}
\usepackage{booktabs}
\usepackage{multirow}
\usepackage{url}
\usepackage{tikz}
\usepackage{pgfplots}
\usepgfplotslibrary{groupplots}
\pgfplotsset{compat=1.18}
\usepackage{subcaption}  
\usepackage{xspace}
\usepackage{setspace}
\usepackage{amsthm}
\newtheorem{definition}{Definition}
\usepackage{listings}
\definecolor{leanBlue}{RGB}{35,75,135}
\definecolor{leanGreen}{RGB}{35,115,75}
\definecolor{leanGray}{RGB}{105,105,105}
\definecolor{leanPurple}{RGB}{115,65,145}
\definecolor{leanBackground}{RGB}{248,248,248}

\lstdefinelanguage{Lean}{
morekeywords={
theorem, lemma, def, example, by, where, let, have,
show, fun, intro, intros, exact, apply, constructor,
cases, induction, simp, rw, unfold, decide, trivial,
namespace, end, section, variable, variables,
import, open, structure, class, instance, deriving,
match, with, if, then, else, do, return, Type, Prop,
true, false
},
morekeywords=[2]{
Nat, Int, Bool, List, Option, String, Array,
And, Or, Not, Exists, Eq
},
sensitive=true,
morecomment=[l]{--},
morecomment=[s]{/-}{-/},
morestring=[b]",
}

\lstdefinestyle{lean}{
language=Lean,
basicstyle=\ttfamily\small,
keywordstyle=\color{leanBlue}\bfseries,
keywordstyle=[2]\color{leanPurple},
commentstyle=\color{leanGray}\itshape,
stringstyle=\color{leanGreen},
backgroundcolor=\color{leanBackground},
frame=single,
rulecolor=\color{black!15},
framerule=0.3pt,
framesep=5pt,
numbers=none,
breaklines=true,
breakatwhitespace=false,
keepspaces=true,
columns=fullflexible,
showstringspaces=false,
tabsize=2,
aboveskip=0.8em,
belowskip=0.8em,
}

\newtheorem{example}{Example}
\usepackage{amsmath}
\usepackage{tabularx}
\usetikzlibrary{calc}
\newif\ifComments

\Commentstrue

\begin{document}
\newcommand{\methodName}{\textsc{AssureCode}\xspace}\newcommand{\old}[1]{\textcolor{gray}{#1}}
\newcommand{\weakDelete}[1]{\textcolor{gray}{#1}}
\newcommand{\nc}[1]{\textcolor{blue}{NC: #1}}

\title{\textsc{AssureCode}: Structured Justification for Trustworthy Agentic Coding}

\author{Logan Murphy}
\affiliation{%
  \institution{University of Toronto}
  \city{Toronto}
  \country{Canada}
}
\email{lmurphy@cs.toronto.edu}

\author{Wilson Sy}
\affiliation{%
  \institution{University of Toronto}
  \city{Toronto}
  \country{Canada}
}
\email{wilsons@cs.toronto.edu}

\author{Rhys Balevicius}
\affiliation{%
  \institution{University of Toronto}
  \city{Toronto}
  \country{Canada}
}
\email{balevicius@cs.toronto.edu}

\author{Kelsie Fung}
\affiliation{%
  \institution{University of Toronto}
  \city{Toronto}
  \country{Canada}
}
\email{kelsiefung@cs.toronto.edu}

\author{Karenna Kung}
\affiliation{%
  \institution{University of Toronto}
  \city{Toronto}
  \country{Canada}
}
\email{kkung@cs.toronto.edu}

\author{Aren A. Babikian}
\affiliation{%
  \institution{McGill University}
  \city{Montreal}
  \country{Canada}
}
\email{aren.babikian@mcgill.ca}

\author{Marsha Chechik}
\affiliation{%
  \institution{University of Toronto}
  \city{Toronto}
  \country{Canada}
}
\email{chechik@cs.toronto.edu}

\renewcommand{\shortauthors}{Murphy et al.}
\begin{abstract} 
The use of AI-based tools in software development is rapidly changing how software is produced, enabling large volumes of code to be generated with limited human implementation effort. However, this raises fundamental challenges for ensuring that generated artifacts are aligned with developer intent and have been appropriately validated. In this work, we propose the use of \emph{structured argumentation} in driving AI-assisted software development. Inspired by techniques from assurance engineering for safety-critical systems, we define a language of \emph{justification rationales} that explicitly relate informal natural language prompts, formal specifications, assumptions about developer intentions, and program analysis evidence in a structured and verifiable format. We then describe \textsc{AssureCode}, a tool-supported methodology for creating such rationales as part of agentic code generation. We provide an empirical analysis of our implementation where we investigate its capacity to contribute towards justified assurance of correctness for AI-generated code.
\end{abstract}

\ccsdesc[500]{Software and its engineering~Software verification and validation}
\ccsdesc[500]{Software and its engineering~Formal methods}

\keywords{Agentic software engineering, trustworthy AI,
assurance cases, formal methods}
\maketitle


%
%


\section{Introduction}
\label{sec:intro}

The software engineering landscape is undergoing a radical transformation through the widespread use of LLM-based (``agentic'') tools as software development assistants~\cite{he2025llm}. These tools allow for large volumes of code to be rapidly generated, which has the potential to  increase developer productivity. However, this scale of generation also raises the question of \emph{trust} in the generated artifacts.  In traditional software development, developers are directly involved in implementing their systems and can draw on their understanding of the development context and requirements to assess whether the resulting software is adequate. AI-assisted software development creates a greater separation between what the developer intends and what gets produced.

Accordingly, the role of human developers is evolving, with less emphasis on directly implementing code and more emphasis on specifying requirements, steering agentic tools, and evaluating their outputs. In order to have trustworthy code, we require a sufficiently clear and precise account of what the software is intended to accomplish (e.g., requirements and specifications), and we must have concrete evidence that the software satisfies particular properties (e.g., successful execution of tests). We must also have grounds for believing that this evidence is sufficient to establish alignment between the software artifact and the developer's intentions. Because of their centrality in establishing trustworthiness, offloading specification of intention and determination of evidence adequacy to agentic tools is {more problematic} than offloading the implementation of code.
An agent may produce an internally consistent set of requirements, implementation, and tests while still failing to demonstrate alignment with what the developer actually intended. We therefore argue that trustworthy agentic development requires a disciplined notion of \emph{justification}: an explicit and systematic account of reasons why the requirements, implementation, and evidence produced during development warrant trusting that the resulting software realizes the developer's intentions. 

To this end, this work introduces a language of structured \emph{justification rationales}, inspired by structured assurance cases developed for safety-critical systems~\cite{kelly2004goal,rushby2015interpretation}. A justification rationale explicates not only the properties a program should satisfy and the evidence used to establish them, but also the additional issues that must be addressed to justify the adequacy of this evidence. We refer to these issues as \emph{collateral validation obligations}~\cite{heimdahl2007safety}.
A justification rationale makes explicit the reasoning connecting claims, evidence, and validation obligations needed to establish alignment between developer intent and generated artifacts. We then introduce \methodName, a tool-supported methodology for constructing and using such rationales as part of agentic code generation, combining autoformalization, theorem proving, and agentic artifact generation.

\emph{Contributions.} This paper makes the following contributions: (1) We define a formalism for \emph{justification rationales} for code generation, analysis, and validation, inspired by assurance cases in safety-critical software engineering~\cite{rushby2015interpretation}.
 (2) We describe \methodName, a tool-supported methodology for agentic software development based on the semi-automated construction and analysis of justification rationales.  
(3) We provide an empirical evaluation of our implementation of \methodName, investigating its ability to identify relevant collateral validation obligations that should be reviewed by human developers before accepting AI-generated code.

\section{Motivation and Background}
\label{sec:background}


\subsection{Trustworthiness in Agentic Software Development}
\label{sec:background:motive}
As outlined in Sec.~\ref{sec:intro}, our starting observation is that in order for software to be \emph{trusted}, at least three elements are needed: (1) \emph{intention}, i.e., what the developers of the artifact intend for it to do (and not do); (2) \emph{verification}, i.e., some analysis or investigation into what properties the software satisfies, or how it behaves under certain conditions, yielding some form of concrete \emph{verification evidence}; and (3) \emph{justification}, which is some form of warrant or reasoning by which developers \emph{believe} that the collected verification evidence demonstrates alignment of the software artifact to their established intention.
Each of these is essential: without a clear intention, one cannot distinguish software suitable for deployment from software that is unsuitable; without verification, one cannot predict what software will do when it is deployed; and without justification, one cannot know that  verification evidence collected actually shows that software is aligned with one's intention.

In traditional (i.e., non-AI) software development, each of these components is present to some extent, though sometimes implicitly; this is especially true for justification. 
This lack of explicit justification is often tolerable, at least in part, because there are human developers who can be held accountable to provide a justification as to how a specification should be interpreted, why certain implementation decisions were made, or why certain forms of analysis were conducted. Agentic software development offers a radically different set of circumstances. 


\paragraph{Motivating Example}
\begin{figure}
    \centering
\includegraphics[width=0.8\linewidth]{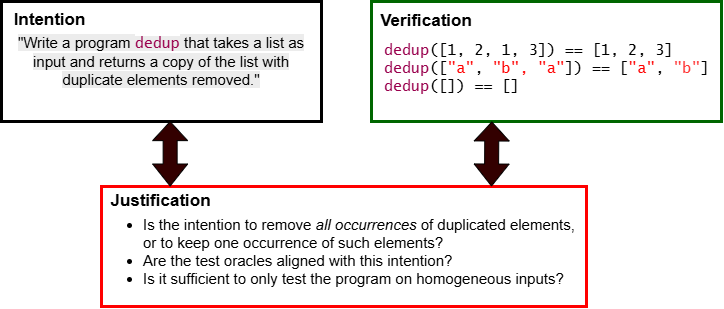}
    \caption{\bf Justification as a bridge between intention and verification for the motivating example.
    }
    \label{fig:motiv_example_trust}
    \vspace{-4mm}
\end{figure}
Consider a simple vibe coding scenario in which a developer prompts an agent to generate a Python program $\textcolor{purple}{\tt dedup()}$ that removes duplicate elements from a list (Fig.~\ref{fig:motiv_example_trust}, top left). As part of the task, the agent also crafts a small test suite (Fig.~\ref{fig:motiv_example_trust}, top right). Supposing the generated code passes the tests, what should be the reason that one is confident that the program is aligned with the developer's intention? As part of this inquiry, one could try to seek clarification about the intention itself: for instance, whether the intent is to remove \emph{all occurrences} of duplicated elements, or merely to keep a single copy of each such element~\cite{endres2024can}; and having made this clarification, one would inspect the test oracles to make sure they are indeed aligned with this intention. One could also scrutinize the diversity of the tests: for instance, if the test suite only considers homogeneous lists, one could ask whether this is sufficient, or whether the program is also expected to handle heterogeneous lists. Through the exercise of answering questions such as these, one can arrive at a more defensible justification for accepting the code (Fig.~\ref{fig:motiv_example_trust}, bottom). 

{These kinds of considerations often emerge naturally in traditional software development, a process which we,} following Heimdahl, refer to as \emph{collateral validation}~\cite{heimdahl2007safety}.   We refer to the questions produced through this inquiry as \emph{collateral validation obligations}, and we view the elicitation and resolution of these obligations to be an essential part of the justification process. In many agentic development settings, scrutiny of this kind is ultimately left to the developer to perform in an \emph{ad hoc} fashion, using their own subjective judgment. This can lead to inconsistency within and between projects, as well as risk the deployment of code that ``looks right'' without having undergone sufficient scrutiny. Moreover, as the volume of generated artifacts and accompanying explanations increases, developers may face substantial cognitive demands in determining what requires scrutiny.
{To mitigate these issues, we seek to make  justification an \emph{explicit} and \emph{systematic} part of agentic development. We find} inspiration in the domain of safety-critical systems, in which this kind of justification is performed as a matter of course.
\subsection{Structured Assurance Cases}
\label{sec:background:ac}
{In safety-critical systems engineering (automotive, aviation, etc.), organizations are expected to justify the safety of their systems through the use of structured \emph{assurance cases}, i.e., structured arguments explaining how a system's safety objectives have been achieved.} Per the automotive safety standard ISO 26262: ``\emph{An argument without supporting evidence is unfounded. Evidence without an argument is unexplained, it can be unclear that, or how, safety objectives have been satisfied}''~\cite{iso26262}.

Various notations and languages for assurance cases have been developed, such as the Goal Structuring Notation (GSN~\cite{kelly2004goal}). GSN assurance cases are tree-like artifacts in which claims about a system, its environment, or its development (called \emph{goals}) are organized into a structured argument, with internal goals being decomposed via \emph{strategies} into subgoals, and leaf goals being supported by evidence artifacts (called \emph{solutions}). Goals and strategies may be augmented with \emph{assumption} nodes (e.g., about the interpretation of a goal), \emph{justification} nodes (explaining the reasoning for a strategy), and \emph{context} nodes (defining terminology or referencing relevant engineering artifacts).

\begin{figure}
    \centering
    \includegraphics[width=0.9\linewidth]{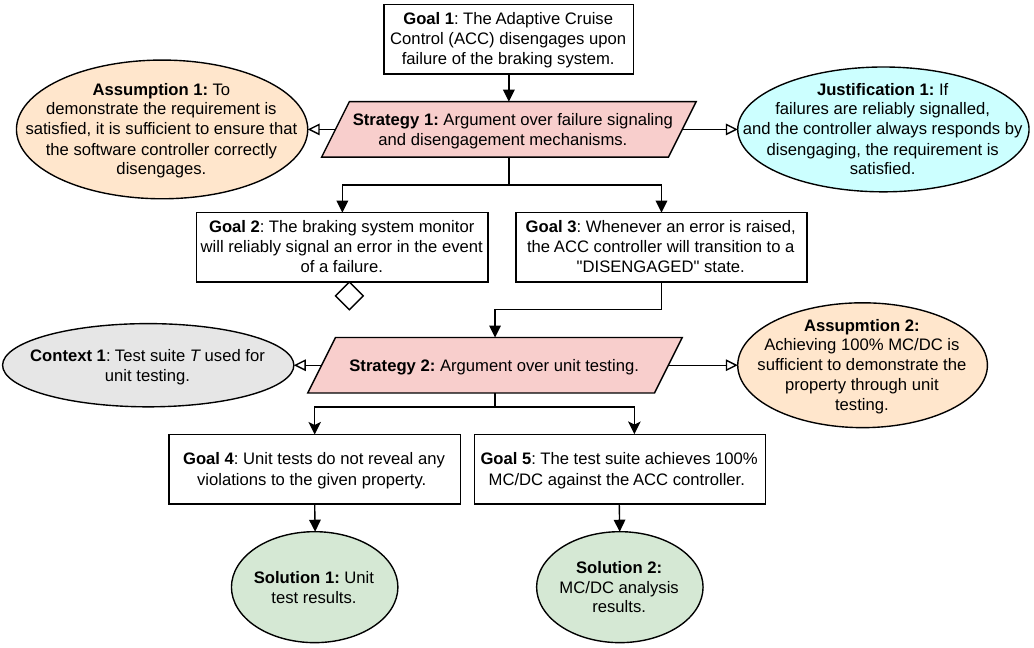}
    \caption{\bf An assurance case fragment for an adaptive cruise control system (adapted from~\cite{viger2026evaluating}). \vspace{-4mm}}
    \label{fig:gsn_example}
\end{figure}

\begin{example} \rm \label{ex:gsn_intro}
    An assurance case fragment for an adaptive cruise control (ACC) system is shown in Fig.~\ref{fig:gsn_example}. The root goal for this fragment (\textbf{Goal 1}) describes a requirement of the ACC to disengage upon a braking failure. This  goal is decomposed by \textbf{Strategy 1} into more precise claims about the failure signalling and disengagement mechanisms, as explained by \textbf{Justification 1}. This strategy is subject to \textbf{Assumption 1}, asserting that it is sufficient to ensure that the ACC software controller demonstrates the desired disengagement behaviour, i.e., the actual effects of the controller disengaging on the vehicle's accelerator hardware are not in scope for this argument. \textbf{Goal 2} is left undeveloped, as indicated by the diamond token, while \textbf{Goal 3} is further decomposed by \textbf{Strategy 2}, yielding an argument over unit testing; the test suite $T$ used is referenced by the  node \textbf{Context 1}. This argument induces one subgoal (\textbf{Goal 4}) which asserts that the test suite did not reveal any violations, and another subgoal (\textbf{Goal 5}) which asserts that the test suite achieves 100\% modified condition/decision coverage (MC/DC) against the controller. Both subgoals are leaves, and can be supported directly by evidence (solutions), whose adequacy is subject to \textbf{Assumption 2}, i.e., that achieving 100\% MC/DC is sufficient to support the given property through unit testing. 
\end{example} 

{As shown in this example, structured assurance cases are well-suited to support structured justification of systems and software adequacy.} The main challenge with using assurance cases in agentic software development is that assurance cases are traditionally developed manually and iteratively, over a long period of time by teams of engineers and domain experts, while agentic software development expects rapid iteration and deployment. Thus, in order to gain \emph{assurance outcomes} in agentic development in a similar fashion to those that are obtained under traditional assurance case development, we need to adapt the assurance case formalism -- and the creation of these artifacts -- to better support agentic software development.

\section{Justification Rationales for Agentic Coding}
\label{sec:rationale}

%


As outlined in the previous section, our goal is to adapt the assurance case formalism to support the systematic and explicit justification of AI-generated code, while reducing the overhead imposed on software developers. In this section, we introduce a formalism for defining \emph{structured justification rationales} (or simply \emph{rationales}) for agentic coding.

Our language retains the core argumentative structure of GSN in its representation of claims, decomposition strategies, assumptions, and evidence. However, unlike typical GSN-style assurance cases, our rationales also need to be amenable to automated construction and analysis; this enables agents to take on as much of the effort of constructing and analyzing the rationale as possible, allowing human developers to focus their attention only where it is truly needed. Any automated analyses should provide strong \emph{guarantees} for certain properties of the rationale (e.g., the soundness of a decomposition or the validity of some evidence) as much as possible. To this end, our language of rationales uses (partial) \emph{formalization} to enable the use of automated reasoning and analysis tools for such guarantees.


\subsection{Justification with (Partial) Formalization}
\label{sec:rat:formalization}
To support automated construction and analysis of justification rationales, we need a formal representation of at least some of the concepts to which the rationale refers. This formalization provides a basis for establishing justification properties mechanically, e.g., whether a set of statements about a program form a logically sound argument. However, defining and validating a complete formalization of the problem space would impose substantial specification and formalization effort on the developer. To mitigate this, we propose to define rationales using a \emph{partial formalization vocabulary} in first-order logic. This vocabulary can be used to formalize only those aspects of the problem space which are straightforward to formalize, or are most essential to support structured argumentation, while allowing other aspects to remain formally uninterpreted. We define such a vocabulary as follows.
\begin{definition}[Vocabulary]
    A \emph{partial formalization vocabulary} $V$ (or simply \emph{vocabulary}) is a pair $(\Sigma, \Gamma)$, where $\Sigma$ (called the \emph{signature}) defines a set of constant symbols, type symbols, function symbols, and predicate symbols, and $\Gamma$ provides a partial theory over $\Sigma$, providing definitions to a subset of the symbols in $\Sigma$.
\label{def:vocabulary}
\end{definition}

\begin{example} \rm \label{ex:motiv_ex_vocab}
Consider again the \textcolor{purple}{\tt dedup} vibe coding scenario described in Sec.~\ref{sec:background:motive}. In defining a vocabulary to (partially) formalize the intention of this prompt, one may define a signature $\Sigma$ which includes symbols for sets corresponding to relevant Python datatypes (e.g., $\mathsf{List}$), functions (e.g., $\mathsf{index}(\ell, i)$, $\mathsf{count}(x,\ell)$), and predicates (e.g., $\mathsf{mem}(x,\ell)$). These are all examples of symbols that would likely be relevant in generating any program involving lists; however, the signature can also define symbols tailored to this problem, e.g., predicates $\mathsf{nodup}:\mathsf{List}\to\mathsf{Prop}$ (i.e., a list contains no duplicates) and $  \mathsf{preservesOrder}:\mathsf{List}\times\mathsf{List}\to\mathsf{Prop}$ (i.e., a list preserves the occurrence order of elements in another list). Given an appropriate signature, one can define a theory $\Gamma$ assigning meaning to (some) of these symbols. Conventional symbols such as $\mathsf{List}$ and $\mathsf{mem()}$ can be given their standard interpretations, while problem-specific predicates such as $\mathsf{nodup}$ and $\mathsf{preservesOrder}$ require more careful formalization. However, we may not need to formalize \emph{both} these predicates right away; in this case, the predicate $\mathsf{preservesOrder}$ does admit a formalization, but it may be rather involved. Thus, $\Gamma$ may initially only define $\mathsf{nodup}(\ell)$ as 
$\mathsf{nodup}(\ell) \triangleq \forall x. \mathsf{mem}(x,\ell) \implies \mathsf{count}(x,\ell) = 1$
while leaving $\mathsf{preservesOrder}$ uninterpreted. Finally, the vocabulary should include a symbol for the \textcolor{purple}{${\tt dedup}$} function itself, which should remain uninterpreted until the program is actually generated.
\end{example}

Having fixed a formalization vocabulary, we can begin to express \emph{claims} to serve as the central elements of the rationale, analogous to goals in GSN. To realize our requirement for the rationale to be both readily intelligible to developers, while also being amenable to automation and analysis, we take rationale claims to be defined by \emph{both} formal and natural language statements. 

\begin{definition}[Claim]\label{def:claim}
    Given a vocabulary $V$, a \emph{claim} over $V$ is a pair $\langle\phi, s\rangle$, where $\phi$ is a formal proposition over $V$, and $s$ is a string expressing $\phi$ in natural language.
\end{definition}

\begin{example} \label{ex:motiv_ex_claim} \rm
    Continuing with the motivating example whose formalization vocabulary was described in Ex.~\ref{ex:motiv_ex_vocab}, one plausible formalization of the prompt as a claim could be $c = \langle\phi, s \rangle$, with 
$ \phi \triangleq \forall \ell. ~(\forall x. ~\mathsf{mem}(x,\ell) \iff \mathsf{mem}(x,\textcolor{purple}{\mathsf{dedup}}(\ell)) ~\wedge \mathsf{nodup}(\textcolor{purple}{\mathsf{dedup}}(\ell)) \wedge \mathsf{preservesOrder}(\ell, \textcolor{purple}{\mathsf{dedup}}(\ell))$
    and $s =$ ``\emph{For every input list, the program returns a list containing exactly the same elements as the input, with no element occurring more than once, and with their initial occurrence order preserved.}'' 
\end{example}


\begin{definition}[Claim Decomposition (adapted from ~\cite{murphy2026assurance})]\label{def:decomp}
Fix a vocabulary $V = (\Sigma,\Gamma).$ A \emph{claim decomposition} is a pair $\langle c, \{c_1,\ldots,c_n\}\rangle$, where $c$ is referred to as the \emph{parent claim}, the $\{c_i\}$ are referred to as the \emph{child claims} or \emph{subclaims}, and all claims are expressed over $V$.  Given $c = \langle\phi,s\rangle$ and each $c_i = \langle\phi_i, s_i\rangle$, the decomposition is said to be \emph{sound} if $(\bigwedge_i \phi_i) \implies \phi$ is a theorem under $\Gamma$.

\end{definition}

\begin{example}\label{ex:motiv_example_decomps} \rm 
 Recall the claim $c = \langle \phi,s\rangle$ defined in Ex.~\ref{ex:motiv_ex_claim}. A simple decomposition which can be defined for this claim can be obtained by defining the subclaims $\{\langle \phi_1,s_1\rangle, \langle \phi_2,s_2\rangle, \langle\phi_3,s_3\rangle\}$, with
\begin{align*}
\phi_1 &\triangleq
\forall \ell : \mathsf{List}.
\forall x.
\mathsf{mem}(x,\ell)
\iff
\mathsf{mem}(x,\textcolor{purple}{\mathsf{dedup}}(\ell)) \qquad
\phi_2 \triangleq
\forall \ell : \mathsf{List}.
\mathsf{nodup}(\textcolor{purple}{\mathsf{dedup}}(\ell))\\
\phi_3 &\triangleq
\forall \ell : \mathsf{List}.
\mathsf{preservesOrder}(\ell,\textcolor{purple}{\mathsf{dedup}}(\ell)).
\end{align*}
and $s_1,s_2,s_3$ being the corresponding natural language expressions. It is easy to see that $\phi_1 \wedge \phi_2 \wedge \phi_3 \implies \phi$ is a theorem under any first-order theory. Alternatively, the claim can be decomposed by partitioning the input space, e.g., distinguishing lists that already contain no duplicates and those that do. To do so, we first extend the vocabulary $V$ to include new sets $\mathsf{List_{Dup}}$ and $\mathsf{List_{NDup}}$, which denote the sets of lists that do, and do not, contain duplicates, respectively. We can then define subclaims $\{(\phi_4,s_4), (\phi_5,s_5)\}$, with $\phi_4$ and $\phi_5$ {being copies of $\phi$ quantified instead over $\mathsf{List_{Dup}}$ and $\mathsf{List_{NDup}}$, respectively.}
and $s_4,s_5$ being the corresponding natural language expressions. It is similarly easy to see that $\phi_4 \wedge \phi_5 \implies \phi$ would hold under the updated $\Gamma$. 
\end{example}

\subsection{A Formal Model of Justification Rationales}
\label{sec:rat:full}We now formally define \emph{justification rationales}, an argument formalism inspired by assurance cases (e.g., GSN~\cite{kelly2004goal}) for arguing that a generated artifact satisfies the developer's intent.

\begin{definition}[Rationale]\label{def:rationale}
Given a vocabulary $V$, a \emph{rationale} $r$ over $V$ is defined inductively as (i) a claim $c$ over $V$, called an \emph{undeveloped} claim;
    (ii) a tuple $\langle c, E, O, \alpha\rangle$, where $c$ is a claim over $V$, $E$ is a set of \emph{evidence artifacts}, $O$ is a set of \emph{obligations}, and $E \cup O \neq \emptyset$, and $\alpha$ is an \emph{assertion} that claim $c$ is adequately supported by evidence $E$ given pending resolution of obligations $O$; or
    (iii) a triple $\langle c, \{r_1,\ldots,r_n\}, J\rangle$, where $c$ is a claim over $V$, $r_1,\ldots,r_n$ are rationales over $V$, and $J$ is a \emph{justification}, which is either (i) a \emph{proof} that the root claims of $\{r_i\}$ induce a sound decomposition of $c$, or
        (ii) a pair $\langle \alpha, O\rangle$, where $\alpha$ is an \emph{assertion} that the decomposition is acceptable and $O$ is a set of obligations.
\end{definition}

\paragraph{Evidence and obligations.}
A leaf $\langle c, E, O, \alpha\rangle$ aims to support claim  $c$ by the evidence artifacts $E$, which may include  test results, proofs, or static analysis, together with the resolution of the obligations $O$. An \emph{obligation} is a condition that must be established before the claim can be accepted on the basis of $E$, and it remains \emph{pending} until resolved by a human. That is, in our model, ``evidence'' is produced by tools and has unambiguous interpretations, whereas \emph{obligations} are resolved by humans and may be more subjective in nature. The assertion $\alpha$ associated to a leaf tuple, asserting that the evidence and obligations are collectively sufficient to support $c$, can be viewed as a special ``built-in'' obligation to be confirmed by a human developer.

\paragraph{Justifications.}
In a \emph{decomposition}, the \emph{parent} claim $c$ is supported by the sub-rationales $r_i$ (the \emph{children}), corresponding to the logical inference $\bigl(\bigwedge_i c_i\bigr) \implies c$ with $c_i = \mathit{root}(r_i)$. The justification $J$ takes one of two forms. If it is a \emph{proof} of the implication, the inference is \emph{deductive}. Otherwise, if we cannot \emph{prove} the inference, it can only be imperfectly justified via informal \emph{decomposition obligations} which need to be confirmed by a responsible authority (e.g., a human developer), alongside an \emph{assertion} that these obligations are adequate to trust that the child claims imply the parent claim.



\begin{example}\label{ex:rationale}
\begin{figure}
    \centering \includegraphics[width=\columnwidth]{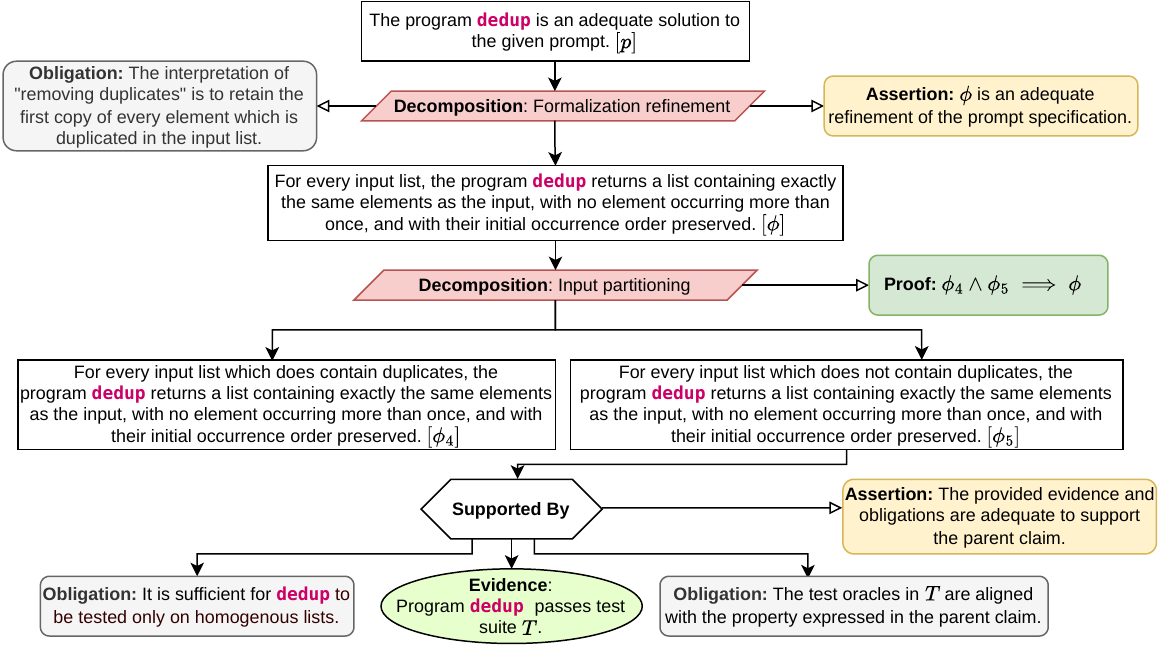}
    \caption{\bf A justification rationale for the running example.}
    \label{fig:rationale_ex}
\end{figure}\rm 
Fig.~\ref{fig:rationale_ex} shows a fragment of a potential rationale for the running example. Claims $\langle \phi,s\rangle$ are written out as ``$s ~[\phi]$'' for legibility. The top-level claim of the rationale asserts that the program $\textcolor{purple}{\mathsf{dedup}}$ is an adequate solution to the user's prompt. Its formal expression is an \emph{uninterpreted} atomic proposition $p$, since the developer's intention as expressed in the prompt has yet to be formalized. The first decomposition layer introduces a formalization refinement; the decomposition attempts to refine the developer's intention as the formal specification $\phi$ defined in Ex.~\ref{ex:motiv_ex_claim}. This decomposition is accompanied by an obligation to confirm that ``removing duplicates'' has been correctly interpreted, and an assertion that, under this assumption, $\phi_1$ is an adequate formalization of the problem. The second layer of decomposition is based on a partition of the input space, as described in Ex.~\ref{ex:motiv_example_decomps}, accompanied by a proof of the logically valid refinement. We only show the full branch for the subclaim asserting correctness for lists which contain duplicates. This claim is supported by one evidence artifact, namely, that the program $\textcolor{purple}{\mathsf{dedup}}$ passes a given test suite $T$ (e.g., the first two tests shown in Fig.~\ref{fig:motiv_example_trust}), accompanied by two obligations: that the oracles in $T$ reflect the property asserted in the parent claim and that it is sufficient for the program to be tested only on homogeneous lists. These are accompanied by an assertion that they are adequate to support the parent claim.

\end{example}

Our justification rationales differ from conventional agentic ``explanations'' by treating justification as a structured argument linking developer intent, implementation, verification evidence, and collateral validation obligations. Natural-language explanations and chain-of-thought~\cite{li2025structured} provide textual accounts of reasoning, but do not explicitly relate claims to implementation artifacts and verification evidence or expose the obligations underlying those relationships.

\section{\methodName: Assurance-Driven Vibe Coding}
\label{sec:overview}
\begin{figure*}
    \centering
\includegraphics[width=\linewidth]{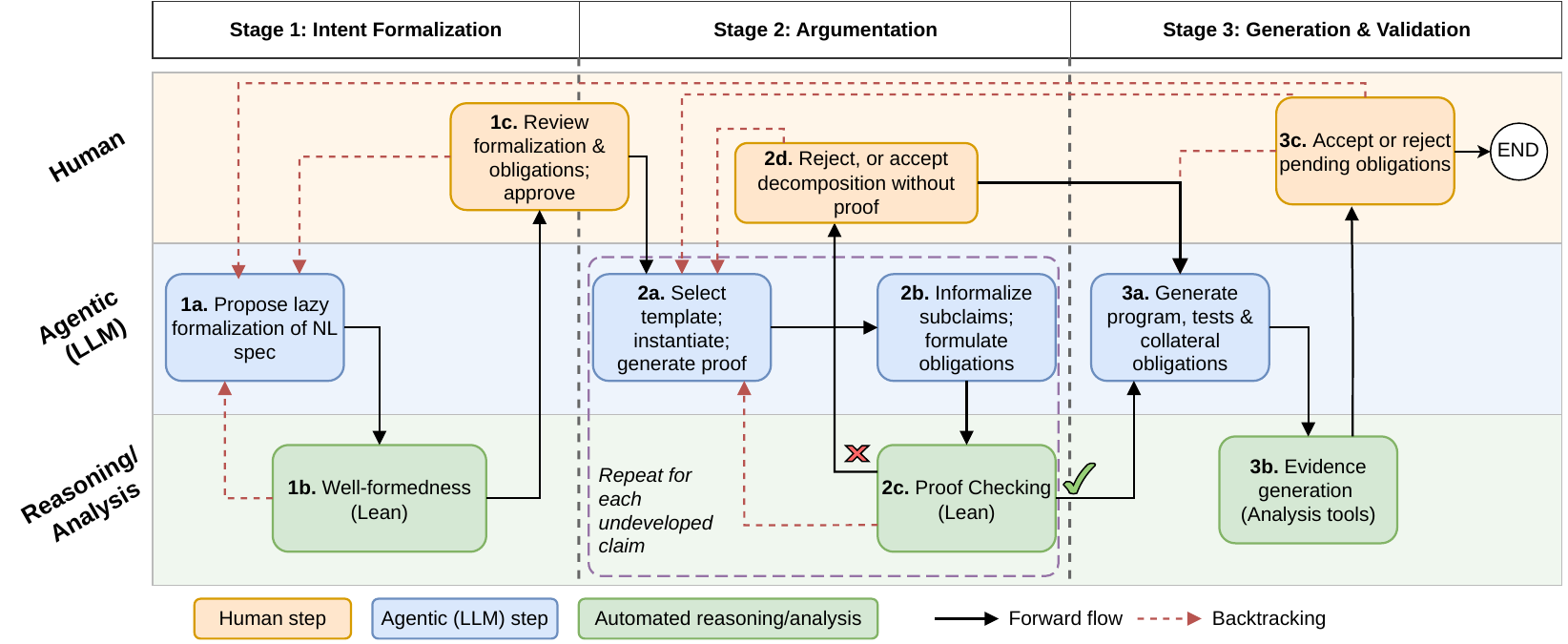}
    \caption{\bf Overview of  the \methodName methodology.}
\label{fig:argos}
\end{figure*}

In this section, we provide an overview of the \methodName development methodology, which aims to support the creation of structured justification rationales as defined in Sec.~\ref{sec:rationale} as part of agentic code generation.  As illustrated in Fig.~\ref{fig:argos}, \methodName consists of three primary and incremental stages, each of which integrates human tasks, agentic tasks performed by LLMs, and automated reasoning and analysis support. {We have developed a prototype of \methodName as a Model Context Protocol (MCP,~\cite{hou2026model}) service. It is configurable by design: the models, automated reasoning support, claim decomposition strategies, and analysis tools are all determined by configurable registries. Our current prototype employs the Lean theorem prover~\cite{de2015lean} as a formalization backend and theorem prover, but our architecture is designed to accommodate alternative tools, e.g., SMT solvers~\cite{barrett2018satisfiability}.} 

\subsection{Stage 1: Intent Formalization}

Intent formalization is the first stage in the workflow, transforming a natural-language specification (i.e., a prompt) into a partially formalized specification suitable for subsequent argumentation, as was done in Ex.~\ref{ex:rationale}. Unlike traditional upfront specification approaches, intent formalization in our setting is  \emph{lazy}.
Given the input spec, the autoformalization agent proposes a formalization consisting of a vocabulary (as defined in Def. \ref{def:vocabulary}) and a functional correctness specification expressed in this vocabulary, both encoded in a formal language supporting mechanized reasoning (in our case, Lean) (Fig.~\ref{fig:argos}-1a). The functional correctness specification uses an opaque identifier for the program (which at this stage, has not yet been generated). This formalization may not be complete or final; some predicates or functions may be uninterpreted. Additional formal interpretation is introduced only when required to support argumentation or verification tasks. 
Lean's syntax and type system establish well-formedness of the formalization (Fig.~\ref{fig:argos}-1b), after which the human reviewer judges whether the formalization captures their intent (Fig.~\ref{fig:argos}-1c). If the formalization is approved by the human user, this formalization refinement is used to initialize the first layer of decomposition of the rationale (as illustrated in Fig.~\ref{fig:rationale_ex}), before we proceed to subsequent argumentation.\footnote{\methodName assumes that users are capable of inspecting and scrutinizing such formalizations directly in Lean. Making this review process more accessible to users without formal methods training is left as future work.}

\begin{example} \rm
Considering again the motivating example, the intent formalization stage uses autoformalization to generate the vocabulary $V$ defined in Ex.~\ref{ex:motiv_ex_vocab}, encoded in Lean. The uninterpreted predicate ${\tt preservesOrder}$ is left as an opaque constant expression. This vocabulary allows us to express the formalized correctness specification as
\begin{lstlisting}[mathescape=true]
def spec : Prop :=
$\forall$ l, ($\forall$ x, x $\in$ l $\leftrightarrow$ x $\in$ (dedup l))  $\land$ nodup (dedup l) $\land$ preservesOrder l (dedup l)    
\end{lstlisting}
which is used as the formal content of the subclaim in the first layer of decomposition (as in Fig.~\ref{fig:rationale_ex}).
    
\end{example}

\subsection{Stage 2: Argumentation}
In this stage, the argumentation agent progressively develops undeveloped claims, ultimately producing a partial rationale populated with explicit verification and validation obligations. These obligations serve as placeholders for evidence that must subsequently be discharged through artifact generation (e.g., code, tests, or proofs) or human validation. Claim decomposition is guided by formal \textit{argument templates} (or simply \textit{templates}), following recent work on automated assurance-case development~\cite{viger2023foremost,murphy2026assurance}. Templates define reusable argumentation patterns together with criteria for determining when their application yields a sound decomposition. Each template specifies (i) the shapes of formal claims it can decompose, implemented via regular-expression matching; (ii) an instantiation schema defining the data required to generate child claims; and (iii) a template for soundness proofs for instantiations of the template, which an agent completes after instantiation.

In our implementation of \methodName, the argumentation agent selects an applicable template from a template \emph{catalogue} and generates the required instantiation data (Fig.~\ref{fig:argos}-2a). Template application operates on the \emph{formal interpretation} of the parent claim: the engine instantiates the corresponding Lean template and extracts the resulting formal subclaims, after which an agent completes the proof of decomposition soundness. Agents are also able to \emph{extend} the formalization vocabulary to facilitate the decomposition and proof. The formal subclaims produced via the template are automatically reinterpreted in natural language via agentic \emph{informalization}~\cite{wu2022autoformalization}, yielding the subclaims presented in the user-facing rationale and used to formulate associated validation obligations (Fig.~\ref{fig:argos}-2b). We do not currently perform a systematic check of semantic alignment between the formal and informalized subclaims; incorporating techniques for checking such alignment~\cite{liu2025rethinking} is left to future work. Next, we attempt to verify the generated proof with Lean (Fig.~\ref{fig:argos}-2c). If the instantiated decomposition cannot be verified, the agent may attempt an alternative instantiation or template, or defer to human intervention. In the latter case, the user's approval is recorded as justification for the inference. Otherwise, we proceed to the next undeveloped claim.

Our current prototype of \methodName provides a base catalogue of four decomposition templates: \emph{domain decomposition}~\cite{murphy2026assurance}, used to partition the input spaces of programs, as in the second decomposition of Ex.~\ref{ex:motiv_example_decomps}; \emph{property decomposition},  which is used to refine a claim of the form $\forall x. P(x)$ by defining new predicates $P_1,\ldots,P_n$ such that we can prove $\forall x. \left(\bigwedge _i P_i(x)\right) \implies P(x)$~\cite{viger2023foremost}; \emph{tautological decomposition}, which supports decompositions based on simple first-order logic tautologies (as in the first decomposition of Ex.~\ref{ex:motiv_example_decomps}, and which uses native proof automation in Lean to avoid relying on an LLM-supplied proof); and \emph{freeform decomposition}, which allows the agent to produce arbitrary subclaims directly, but which also requires the agent to produce a proof from scratch.

\begin{example}\rm The decomposition based on input partitioning in Fig.~\ref{fig:rationale_ex} can be achieved via the \emph{domain decomposition} template. The agent performing the instantiation can first extend the formalization vocabulary with the two domains of inputs, defined as sets of lists:
\begin{lstlisting}[mathescape=true]
def ListDup := {l | $\neg$ l.Nodup}  
def ListNDup  := {l | l.Nodup}
\end{lstlisting}
The domain decomposition is then instantiated by the agent by specifying a family of sets of lists:
\begin{lstlisting}
def classes : Family := {ListDup, ListNDup}
\end{lstlisting}
To establish that the resulting subclaims soundly decompose the parent claim, the template initializes a soundness proof, which is completed by the agent and verified by Lean:
\begin{lstlisting}
theorem completeness : complete classes Set.univ := 
 by intros l h; by_cases h : Nodup l
    . use ListNDup ; rw [ListNDup,classes] ; simp [h]
    . use ListDup ; rw [ListDup, classes] ; simp [h]
\end{lstlisting}
\end{example}

In general, argumentation may proceed iteratively until the user determines the claims have been sufficiently refined to warrant support by evidence, or the argumentation agent can make this determination; both configurations are supported by \methodName. Alternatively, the argumentation process may only extend to a certain depth specified by the user at the start of the session, as is the default configuration of our prototype. Once argumentation has completed, we are left with a \emph{partial} rationale whose leaf claims are undeveloped. We next transition to the \emph{generation} stage.

\subsection{Stage 3: Generation}

{Having developed a partial rationale, \methodName{} proceeds to generating evidence and obligations to support leaf claims using  structured patterns(Fig.~\ref{fig:argos}-3a). Following~\cite{murphy2026assurance}, we refer to these patterns as \emph{analytic} templates, distinct from the \emph{logical} templates used for argumentation (Stage~2). In the default configuration of \methodName, a \emph{functional testing template} is used to generate a test suite together with collateral validation obligations concerning its adequacy and underlying assumptions. These artifacts and obligations are then organized into a structured workflow of generation, analysis, and manual validation tasks.} {\methodName{} then automatically executes this workflow, generating relevant artifacts, invoking available analysis tools (Fig.~\ref{fig:argos}-3b), and referring obligations to the human user (Fig.~\ref{fig:argos}-3c). For example, under the testing template, because the test oracle is associated with the corresponding formal claim in Lean, we can partially automate proof of their alignment when the formalization provides a sufficiently interpreted vocabulary and the alignment falls within our available proof automation. When such a proof succeeds, it is incorporated as evidence in the rationale; otherwise, the alignment remains an explicit human obligation, as illustrated in Fig.~\ref{fig:rationale_ex}. For instance,
{mirroring the bottom layer of the rationale in Fig.~\ref{fig:rationale_ex}, the testing template induces a workflow that generates and analyses the test suite before executing it against the generated implementation. The workflow first generates the test suite and verifies its alignment with the functional specification, after which the user reviews its adequacy before the tests are executed against the generated implementation. The precise ordering of these activities is a property of the analytic template and can be reconfigured; e.g., the template could defer human review until after test execution.} Once all goals have been supported by appropriate evidence, with obligations approved by the user, the workflow terminates, resulting in a fully developed and approved rationale for the generated program.

\newcommand{\toadd}[1]{\ifComments
\textcolor{green}{\textit{TODO: #1}}
\fi}
\newcommand{\todo}[1]{\toadd{#1}}
\section{Empirical Evaluation}
\label{sec:evaluation}
In this section, we evaluate the degree to which \methodName{} can support the development of \emph{assurance of correctness}, comparing it with a practical baseline in which a developer uses an LLM to assist with assurance activities. Specifically, we consider a baseline process in which a developer iteratively uses an LLM to identify potential concerns and validation obligations that should be considered before accepting an AI-generated program as adequate on the basis of having passed a given test suite. We refer to this process as \emph{vibe justification}. The baseline is intended to represent a plausible practical alternative to \methodName{}, as it reflects the use of LLMs to support assurance activities without imposing a prescribed methodology for organizing the associated artifacts. Thus, we frame our evaluation around the following research questions:

\begin{tcolorbox}[
    colback=gray!5,
    colframe=gray!40,
    boxrule=0.4pt,
    arc=1pt,
    left=4pt,
    right=4pt,
    top=3pt,
    bottom=3pt
]\small 
\begin{description}[leftmargin=*,labelsep=0.5em]
    \item[\textbf{RQ1 (Volume).}] How many collateral validation obligations
    does each approach generate?

    \item[\textbf{RQ2 (Coverage).}] To what extent does each approach
    address obligations produced by the other?

    \item[\textbf{RQ3 (Coverage basis).}] To what extent does partial 
    formalization address obligations from the
    baseline?

    \item[\textbf{RQ4 (Assurance value).}] What is the assurance
    value of concerns identified by only one approach?
\end{description}
\end{tcolorbox}

These questions centre the evaluation on collateral validation obligations because they provide the connective tissue between  sources of assurance: the obligations themselves, together with the associated formalization, reasoning, and verification artifacts. Importantly, obligations generated by \methodName{} cannot be interpreted independently of these artifacts, as they are defined relative to what other artifacts already establish; for example, an ambiguity resolved by an approved formalization need not give rise to a separate validation obligation. Accordingly, our coverage analysis considers not only obligation-level correspondence but also, for \methodName{}, the contribution of associated formalization artifacts. We do not separately evaluate the individual stages of \methodName{}, such as autoformalization~\cite{wu2022autoformalization}, agentic theorem proving~\cite{jiang2022draft}, agentic test generation~\cite{chen2024chatunitest}, and code generation~\cite{mathews2024test}. These constituent techniques have been, and continue to be, studied separately; our focus is on how artifacts generated by their integration through \methodName{} collectively contribute to the assurance of correctness. Finally, our goal is not necessarily for \methodName{} to significantly outperform vibe justification on every measure. Rather, because \methodName{} provides a more structured and systematic process, evidence that it can achieve comparable degrees of assurance would support its credibility as an alternative to vibe justification.

\subsection{Evaluation Methodology}
\label{sec:evaluation-methodology}

We now describe the components of our evaluation process.\footnote{Experiments were executed on a standard laptop ({10-core Intel i5-13450HX CPU with 16GB RAM).} Our implementation, results and analyses, is available online \href{https://doi.org/10.5281/zenodo.23114525}{https://doi.org/10.5281/zenodo.23114525}}

\subsubsection{Evaluated benchmarks}
We evaluated \methodName{} and the vibe justification baseline on a set of prompt-based Python code generation tasks drawn from three established code-generation benchmarks:
HumanEval, a widely used benchmark consisting of hand-written Python function-completion problems~\cite{chen2021evaluating};
LiveCodeBench, which draws on recent, timestamped contest problems from LeetCode, AtCoder, and Codeforces~\cite{jain2025livecodebench};
and MBPP, which provides crowd-sourced, entry-level Python problems~\cite{austin2021program}.
For all benchmarks, each task includes a representation of the programming problem, expressed in English, and a test suite used to assess the correctness of a candidate solution. Because our evaluation requires manual assessment of generated obligations and their potential assurance value, we randomly sampled ten problems from each benchmark, giving a corpus of 30 code-generation tasks. This sample size enables close inspection of the obligations and associated artifacts produced during the evaluation while keeping manual analysis feasible.
\subsubsection{Subject LLMs}
\label{sec:compared-llms}
We evaluate \methodName{} using four contemporary general-purpose LLMs from several major providers: {\tt gpt-5.6-sol} (OpenAI), {\tt claude-fable-5.1} (Anthropic),\\ {\tt deepseek-v4-pro} (DeepSeek), and {\tt gemini-3.1-pro} (Google). This selection provides diversity across model families and provider ecosystems, while also including an open-weight model ({\tt deepseek-v4-pro}) alongside proprietary models. For each LLM, we instantiate both the vibe justification baseline and \methodName{} using the same underlying model, enabling a direct comparison of the resulting assurance artifacts and collateral validation obligations. This yields 12 benchmark--LLM configurations, and, given our corpus of 10 code-generation tasks per benchmark, results in $ 120$ paired evaluations.

\subsubsection{Baseline: Eliciting Obligations via ``Vibe Justification''}
\label{sec:baseline-obligation-generation} 

At a high level, a prompt for vibe justification amounts to saying the following: ``\emph{Consider specification $S$ and test suite $T$. What concerns, inquiries, or questions should a diligent software developer consider before accepting a program $P$ as an adequate solution to $S$ on the basis of having passed all tests in $T$?}''. In preliminary explorations of ``vibe justification'', we experimented with two styles which further elaborate the instructions for the LLM agent: a \emph{conservative} style that instructs the agent to be careful in proposing an obligation only when there is a good reason to suspect potential misalignment between intention, specification, and test artifacts. The \emph{liberal} style adopts a lower threshold for raising potential concerns. These preliminary explorations found that, while these strategies produced overlapping sets of obligations, the results of one rarely subsume those of the other. Thus, to generate a diverse set of obligations for our baseline, we employed \emph{both} styles of prompt in our baseline generation.

\emph{Iteration.}
Given the stochastic nature of LLMs, and to generate a diverse set of obligations for our baseline, we queried each of our subject LLMs 3 times with both the conservative and liberal prompt. Consequently, across the 120 trials, we obtain \(2\cdot 3 \cdot 120 = 720\) sets of validation obligations. 

\emph{Deduplication.} 
Having generated 6 sets of obligations for each trial, we applied a multi-stage deduplication process to obtain sets of unique obligations. First, we embed each obligation using \texttt{BAAI/bge-large-en-v1.5}, an encoder model based on BERT~\cite{devlin2019bert}, and compare the resulting vectors using cosine similarity, a common technique for measuring semantic similarity in natural-language processing~\cite{wang2020measurement}. Next, we deduplicate obligations within each generated set for a single iteration by clustering them according to their pairwise similarities. We use complete-linkage clustering at a cosine threshold of 0.85 for baseline obligations: every pair in a cluster must meet the threshold. Each cluster retains its original obligations and selects a medoid member as its representative. Candidate duplicate pairs identified through embedding-based comparison are also assessed by an LLM judge. We then deduplicate across the \(k\) sets generated for a fixed task, LLM, and obligation-generation strategy, thereby aggregating obligations across iterations. We then deduplicate across prompting strategies, yielding one aggregated obligation set for each task and LLM. For \methodName obligations, we use the same embedding model and complete-linkage procedure with a cosine threshold of 0.9. {Both thresholds were chosen based on preliminary experiments; the difference between the thresholds is presumably due to the obligations generated by \methodName being expressed in a somewhat different tone of language than those generated by the baseline.} Finally, we manually review the resulting clusters to identify any outstanding duplicates or revert merges.

\subsubsection{Eliciting Obligations via \methodName}
For each task and subject LLM, we run \methodName{} once on the task prompt. During these runs, the subject LLM performs all agentic actions in the workflow (formalization, decomposition, verification, and generation), and the resulting rationale tree is constrained to the three-layer structure shown in Fig.~\ref{fig:rationale_ex}. We automatically approve review requests so that each run can proceed without human intervention during execution. {This design choice  allows us to evaluate the obligations and associated reasoning artifacts produced by \methodName{} under a fixed and reproducible execution protocol, without introducing additional variability from intermediate human  decisions.}
The execution of \methodName{} produces collateral validation obligations in the context of the generated assurance artifacts. The initial formalization creates human obligations concerning its fidelity to the task specification; logical argumentation templates induce human obligations concerning that decomposition before Lean checks the proof; and the test generation procedure likewise generates evidence adequacy obligations. We extract all of these,  alongside their associated formalization artifacts.

\newcommand{\boxedVal}[2]{%
\noindent
\begin{tabularx}{\linewidth}{|X|}\hline
#1 
#2\\\hline
\end{tabularx}}
\newcommand{\ranswer}[2]{\boxedVal{\textbf{RQ#1}}{\emph{#2}}}

\subsubsection{Comparing the Baseline and \methodName Artifacts}
\label{sec:baseline-v-method}
{To answer \textbf{RQ2}, we assess mutual coverage between the generated obligation sets by asking whether the artifacts produced by one approach would resolve the validation obligations raised by the other approach. We classify coverage as \emph{full}, \emph{partial}, or \emph{none}. A partial coverage judgment is accompanied by a  residual concern (e.g., a particular kind of edge case missing from a test suite). Our coverage assessment allows multiple obligations to collectively cover one obligation or one obligation to cover multiple obligations. Because \methodName{} produces obligations in the context of additional assurance artifacts, coverage is assessed asymmetrically: baseline obligations can be covered by a combination of obligations and their associated formalization artifacts produced by \methodName{}, whereas \methodName{} obligations can only be covered by obligations produced by the baseline.}

{Finally, to answer \textbf{RQ4}, we assess the potential assurance value of concerns identified by only one approach. For obligations that the other approach does not fully address in RQ2, we  first identify the \emph{concern} that remains unaddressed: the aspect of the specification, test, or justification process that the obligation raises and the other approach does not account for. Obligations that raise the same underlying concern are merged, such that a single concern may correspond to multiple obligations. Each concern is then classified as ``Strong'', ``Fair'', or ``Weak'' according to the degree of assurance value it provides. In particular, we consider whether the concern identifies a substantive issue that a careful developer could otherwise overlook (Strong), provides a useful but limited or partially redundant assurance consideration (Fair), or is obvious, tautological, overly broad, or otherwise unlikely to affect the developer's assessment of correctness (Weak).} 

{For \textbf{RQ2}--\textbf{RQ4}, we use the same LLM judge ({\tt gpt-5.6-sol}) to produce an assessments based on our evaluation criteria for each \textbf{RQ}. The judging prompts were iteratively refined before the full experiment to ensure conformance with these criteria. We additionally conducted a \emph{post hoc} review of 307 randomly sampled decisions: 102 for \textbf{RQ2}, 101 for \textbf{RQ3}, and 104 for \textbf{RQ4}. Alignment with the intended criteria was 96\% for \textbf{RQ2}, 100\% for \textbf{RQ3}, and 84\% for \textbf{RQ4}. The lower alignment for \textbf{RQ4} was consistently conservative, with some concerns classified as Weak'' rather than ``Fair'' and ``Fair'' rather than Strong''. We determined that these results were acceptable, so long as this tendency is kept in mind when interpreting the \textbf{RQ4} results.}

\subsection{Results}

\subsubsection{RQ1}
Table~\ref{tab:cluster-counts} reports the number of validation obligations produced for each of the 12 benchmark--LLM configurations. Overall, \methodName{} produces a comparable number of obligations to the baseline, with averages per configuration ranging from 4.4 to 10.5 obligations for \methodName{} and from 5.0 to 9.4 for the baseline. \methodName{} produces more obligations in 8 of the 12 configurations (e.g., 5.8 versus 8.7 for {\tt deepseek-v4-pro} on HumanEval), fewer in 3 configurations ({\tt gpt-5.6-sol} on HumanEval and LiveCodeBench, and {\tt gemini-3.1-pro} on LiveCodeBench), and approximately the same number in one configuration ({\tt gpt-5.6-sol} on MBPP).
\begin{tcolorbox}[colback=gray!5,
  colframe=gray!40,
  boxrule=0.4pt,
  before skip=4pt,
  after skip=4pt,
  left=5pt,
  right=5pt,
  top=3pt,
  bottom=3pt]
\textbf{Answer to RQ1.}
\methodName{} generates a comparable number of collateral validation obligations to the baseline, with variation depending on the LLM and benchmark. Across the 12 configurations, neither approach consistently produces more obligations than the other, but \methodName generates a larger volume of obligations more frequently.
\end{tcolorbox}

\begin{table}[t]
\centering
\scriptsize
\setlength{\tabcolsep}{3pt}
\begin{tabular}{ll cccc}
\toprule
 & & gpt-5.6-sol & claude-fable-5.1 & deepseek-v4-pro & gemini-3.1-pro \\
\midrule
\multirow{2}{*}{HumanEval}
 & Baseline & \textbf{7.5} (5--11) & \textbf{7.4} (5--9) & \textbf{5.8} (3--8) & \textbf{5.1} (3--7) \\
 & \methodName & \textbf{4.4} (3--6) & \textbf{8.8} (6--14) & \textbf{8.7} (6--11) & \textbf{5.8} (4--7) \\
\midrule
\multirow{2}{*}{LiveCodeBench}
 & Baseline & \textbf{7.1} (4--11) & \textbf{8.3} (6--10) & \textbf{8.5} (5--12) & \textbf{7.3} (6--9) \\
 & \methodName & \textbf{5.1} (4--8) & \textbf{8.6} (5--13) & \textbf{10.5} (7--14) & \textbf{5.5} (4--9) \\
\midrule
\multirow{2}{*}{MBPP}
 & Baseline & \textbf{5.0} (4--7) & \textbf{7.9} (5--11) & \textbf{9.4} (7--11) & \textbf{5.8} (2--9) \\
 & \methodName & \textbf{5.0} (3--10) & \textbf{8.5} (6--11) & \textbf{10.2} (7--24) & \textbf{6.6} (3--13) \\
\bottomrule
\end{tabular}
\caption{\bf Avg. counts (min--max) of obligations generated by the baseline and \methodName per task.}
\label{tab:cluster-counts}
\end{table}
\subsubsection{RQ2}
Fig.~\ref{fig:rq2_coverage}(a) shows how well \methodName{} covers obligations produced by the baseline using the three coverage categories (full, partial, none). Full coverage ranges from roughly 72\% to 88\%, exceeding 80\% in 9 of the 12 configurations. Including partial coverage, \methodName{} addresses at least 90\% of the baseline obligation for all configurations. Coverage is strongest for {\tt gpt-5.6-sol} and {\tt claude-fable-5.1}: their share of uncovered obligations remains low (between 0\% and 5\%) across benchmarks, while for {\tt deepseek-v4-pro} and {\tt gemini-3.1-pro}, between 4\% and 10\% of obligations remain uncovered.
We attribute this result to the strength of the underlying LLM rather than the obligation volume, as we observe no relationship between coverage and obligation counts.

Fig.~\ref{fig:rq2_coverage}(b) shows the reverse direction: coverage of \methodName{} obligations by the baseline. Coverage is consistently lower, with full coverage ranging from roughly 33\% to 80\% and reaching 70\% or more in only 3 of the 12 configurations. The difference is most pronounced for {\tt deepseek-v4-pro} and {\tt gemini-3.1-pro}, where 20--33\% of \methodName{} obligations on HumanEval and LiveCodeBench are not covered by the baseline. This difference is again not simply due to obligation volume; for example, on LiveCodeBench, {\tt gemini-3.1-pro} produces fewer obligations under \methodName{} than under the baseline, while a substantial fraction remain uncovered. 
\begin{tcolorbox}[colback=gray!5,
  colframe=gray!40,
  boxrule=0.4pt,
  before skip=4pt,
  after skip=4pt,
  left=5pt,
  right=5pt,
  top=3pt,
  bottom=3pt]
\textbf{Answer to RQ2.}
\methodName{} provides substantial coverage of baseline obligations, with full coverage exceeding 80\% in most configurations and at least partial coverage in approximately 90\% or more of cases. Coverage of \methodName{} obligations by the baseline is lower, indicating that \methodName{} identifies more concerns not raised by the baseline than vice versa.
\end{tcolorbox}

\subsubsection{RQ3}

\begin{figure}[t]
\centering
\begin{tikzpicture}
\begin{groupplot}[
  group style={
  group name=cov,
  group size=1 by 2,
  vertical sep=0.6cm,
  x descriptions at=edge bottom,
},
  ybar stacked,
  width=0.8\linewidth, height=3.2cm,
  ymin=0, ymax=100, ylabel={\%},
  xtick={0,1,2,3,5,6,7,8,10,11,12,13},
  xticklabels={gpt-5.6-sol,claude-fable-5.1,deepseek-v4-pro,gemini-3.1-pro,gpt-5.6-sol,claude-fable-5.1,deepseek-v4-pro,gemini-3.1-pro,gpt-5.6-sol,claude-fable-5.1,deepseek-v4-pro,gemini-3.1-pro},
  x tick label style={rotate=30, anchor=east, font=\footnotesize},
  xmin=-0.7, xmax=13.7,
  enlarge x limits=false,
  ymajorgrids, tick align=outside,
  title style={font=\footnotesize, yshift=-1ex},
]

\nextgroupplot[
  title={(a) Baseline obligations covered by \methodName},
  legend style={at={(1,1.4)}, anchor=south east, legend columns=-1, font=\footnotesize},
]
  \addplot+[fill=green!60!black, draw=black!70] table[x=x, y=y, row sep=\\] {
      x y \\
      0 88.0 \\
      1 81.1 \\
      2 86.2 \\
      3 88.2 \\
      5 84.5 \\
      6 72.3 \\
      7 77.6 \\
      8 76.7 \\
      10 82.0 \\
      11 83.5 \\
      12 88.3 \\
      13 87.9 \\
  };
  \addlegendentry{FULL}
  \addplot+[fill=orange!80!yellow, draw=black!70] table[x=x, y=y, row sep=\\] {
      x y \\
      0 9.3 \\
      1 16.2 \\
      2 8.6 \\
      3 7.8 \\
      5 12.7 \\
      6 26.5 \\
      7 15.3 \\
      8 13.7 \\
      10 18.0 \\
      11 11.4 \\
      12 3.2 \\
      13 6.9 \\
  };
  \addlegendentry{PARTIAL}
  \addplot+[fill=red!70, draw=black!70] table[x=x, y=y, row sep=\\] {
      x y \\
      0 2.7 \\
      1 2.7 \\
      2 5.2 \\
      3 3.9 \\
      5 2.8 \\
      6 1.2 \\
      7 7.1 \\
      8 9.6 \\
      10 0.0 \\
      11 5.1 \\
      12 8.5 \\
      13 5.2 \\
  };
  \addlegendentry{NONE}
\node[anchor=south, font=\bfseries\small] at (rel axis cs:0.13,1) [yshift=0.45cm] {HumanEval};
\node[anchor=south, font=\bfseries\small] at (rel axis cs:0.50,1) [yshift=0.45cm] {LiveCodeBench};
\node[anchor=south, font=\bfseries\small] at (rel axis cs:0.86,1) [yshift=0.45cm] {MBPP};
\nextgroupplot[title={(b) \methodName obligations covered by baseline}]
  \addplot+[fill=green!60!black, draw=black!70] table[x=x, y=y, row sep=\\] {
      x y \\
      0 47.7 \\
      1 71.6 \\
      2 46.0 \\
      3 36.2 \\
      5 66.7 \\
      6 80.2 \\
      7 61.0 \\
      8 32.7 \\
      10 64.0 \\
      11 72.9 \\
      12 68.6 \\
      13 66.7 \\
  };
  \addplot+[fill=orange!80!yellow, draw=black!70] table[x=x, y=y, row sep=\\] {
      x y \\
      0 45.5 \\
      1 22.7 \\
      2 34.5 \\
      3 41.4 \\
      5 31.4 \\
      6 17.4 \\
      7 16.2 \\
      8 34.5 \\
      10 36.0 \\
      11 22.4 \\
      12 20.6 \\
      13 25.8 \\
  };
  \addplot+[fill=red!70, draw=black!70] table[x=x, y=y, row sep=\\] {
      x y \\
      0 6.8 \\
      1 5.7 \\
      2 19.5 \\
      3 22.4 \\
      5 2.0 \\
      6 2.3 \\
      7 22.9 \\
      8 32.7 \\
      10 0.0 \\
      11 4.7 \\
      12 10.8 \\
      13 7.6 \\
  };
\end{groupplot}
\foreach \x/\name in {0.153/HumanEval, 0.5/LiveCodeBench, 0.847/MBPP} {
  \node[anchor=north, font=\bfseries\small]
    at ($(cov c1r2.south west)!\x!(cov c1r2.south east)+(0,-1.5cm)$) {\name};
}
\end{tikzpicture}
\caption{\bf Obligation coverage (\%) between \methodName{} and the baseline (RQ1).}
\label{fig:rq2_coverage}
\end{figure}
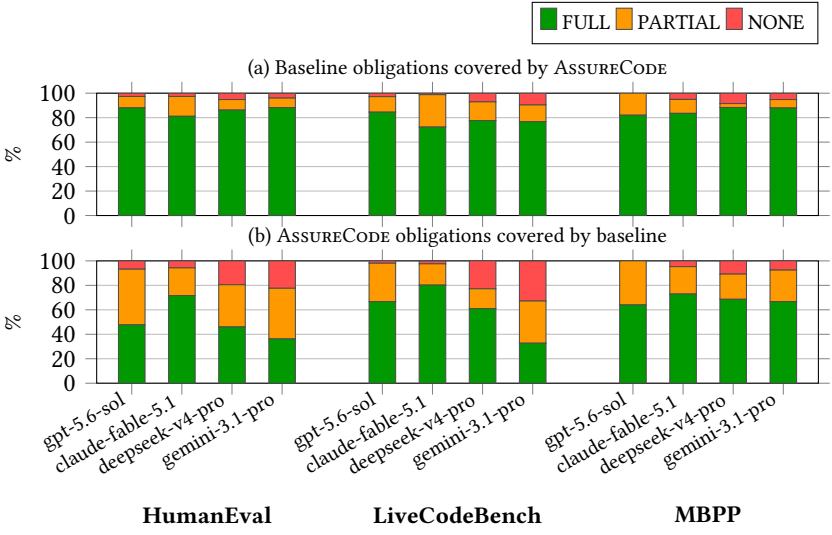

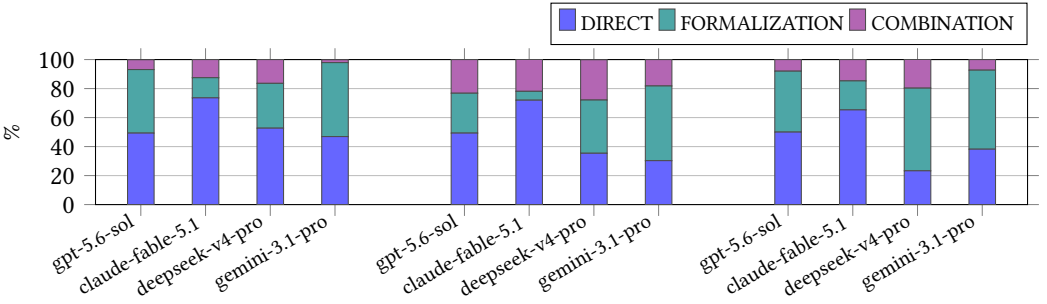
\begin{figure}[t]
\centering
\begin{tikzpicture}
\begin{axis}[
  ybar stacked,
  width=\linewidth, height=3.5cm,
  ymin=0, ymax=100, ylabel={\%},
  xtick={0,1,2,3,5,6,7,8,10,11,12,13},
  xticklabels={gpt-5.6-sol,claude-fable-5.1,deepseek-v4-pro,gemini-3.1-pro,gpt-5.6-sol,claude-fable-5.1,deepseek-v4-pro,gemini-3.1-pro,gpt-5.6-sol,claude-fable-5.1,deepseek-v4-pro,gemini-3.1-pro},
  x tick label style={rotate=30, anchor=east, font=\footnotesize},
  xmin=-0.7, xmax=13.7,
  enlarge x limits=false,
  ymajorgrids, tick align=outside,
  legend style={at={(1,1.1)}, anchor=south east, legend columns=-1, font=\footnotesize},,
]
  \addplot+[fill=blue!60, draw=black!70] table[x=x, y=y, row sep=\\] {
      x y \\
      0 49.3 \\
      1 73.6 \\
      2 52.7 \\
      3 46.9 \\
      5 49.3 \\
      6 72.0 \\
      7 35.4 \\
      8 30.3 \\
      10 50.0 \\
      11 65.3 \\
      12 23.3 \\
      13 38.2 \\
  };
  \addlegendentry{DIRECT}
  \addplot+[fill=teal!70, draw=black!70] table[x=x, y=y, row sep=\\] {
      x y \\
      0 43.8 \\
      1 13.9 \\
      2 30.9 \\
      3 51.0 \\
      5 27.5 \\
      6 6.1 \\
      7 36.7 \\
      8 51.5 \\
      10 42.0 \\
      11 20.0 \\
      12 57.0 \\
      13 54.5 \\
  };
  \addlegendentry{FORMALIZATION}
  \addplot+[fill=violet!60, draw=black!70] table[x=x, y=y, row sep=\\] {
      x y \\
      0 6.8 \\
      1 12.5 \\
      2 16.4 \\
      3 2.0 \\
      5 23.2 \\
      6 22.0 \\
      7 27.8 \\
      8 18.2 \\
      10 8.0 \\
      11 14.7 \\
      12 19.8 \\
      13 7.3 \\
  };
  \addlegendentry{COMBINATION}
  \node[anchor=north, font=\bfseries\small] at (axis cs:1.5,0) [yshift=-1.5cm] {HumanEval};
  \node[anchor=north, font=\bfseries\small] at (axis cs:6.5,0) [yshift=-1.5cm] {LiveCodeBench};
  \node[anchor=north, font=\bfseries\small] at (axis cs:11.5,0) [yshift=-1.5cm] {MBPP};
\end{axis}
\end{tikzpicture}
\caption{\bf Bases of obligation resolution by \methodName{} against the baseline (RQ3).}
\label{fig:resolution}
\end{figure}

To answer \textbf{RQ3}, Fig.~\ref{fig:resolution} breaks down how baseline obligations covered by \methodName{} are resolved: directly by an \methodName{} obligation, by the \methodName{} formalization alone, or through a combination of the two. The results show that formalization provides a substantial contribution to coverage:  formalization contributes to a majority of covered obligations in 8 of the 12 configurations, either alone or in combination with validation obligations. This is, to a degree, expected: the purpose of formalization is in part to reduce ambiguity in the task specification. When excluding formalization-only resolution, the difference between the two approaches is smaller.
\begin{tcolorbox}[colback=gray!5,
  colframe=gray!40,
  boxrule=0.4pt,
  before skip=4pt,
  after skip=4pt,
  left=5pt,
  right=5pt,
  top=3pt,
  bottom=3pt]
\textbf{Answer to RQ3.}
The coverage achieved by \methodName{} relies substantially on its formalization artifacts: the formalization typically contributes to roughly half, or a majority, of covered baseline obligations, either alone or together with \methodName{} obligations. Direct obligation-based coverage remains common, showing that formalization and collateral validation obligations both contribute to assurance.
\end{tcolorbox}
\subsubsection{RQ4}
\begin{figure}[t]
\centering
\begin{tikzpicture}
\begin{groupplot}[
  group style={
    group name=cov,
    group size=1 by 2,
    vertical sep=0.6cm,
    x descriptions at=edge bottom,
  },
  ybar stacked,
  width=0.92\linewidth, height=3.2cm,
  ymin=0, ymax=100,
  xtick={0,1,2,3,5,6,7,8,10,11,12,13},
  xticklabels={gpt-5.6-sol,claude-fable-5.1,deepseek-v4-pro,gemini-3.1-pro,gpt-5.6-sol,claude-fable-5.1,deepseek-v4-pro,gemini-3.1-pro,gpt-5.6-sol,claude-fable-5.1,deepseek-v4-pro,gemini-3.1-pro},
  x tick label style={rotate=30, anchor=east, font=\footnotesize},
  xmin=-0.7, xmax=13.7,
  enlarge x limits=false,
  ymajorgrids, tick align=outside,
]

\nextgroupplot[
  title={(a) Rankings of concerns raised by the baseline obligations but not covered by \methodName.},
  title style={font=\footnotesize, anchor=south west, at={(0,1)}, yshift=-0.5ex},
  ylabel={\%},
  legend style={at={(1,1.5)}, anchor=south east, legend columns=-1, font=\footnotesize},
]
  \addplot+[fill=blue!65!black, draw=black!70] coordinates {
    (0,37.5) (1,50.0) (2,37.5) (3,50.0) (5,90.0) (6,47.8) (7,52.6) (8,41.2) (10,33.3) (11,27.3) (12,18.2) (13,33.3)};
  \addlegendentry{STRONG}
  \addplot+[fill=blue!30, draw=black!70] coordinates {
    (0,62.5) (1,33.3) (2,62.5) (3,33.3) (5,10.0) (6,52.2) (7,31.6) (8,47.1) (10,55.6) (11,45.5) (12,27.3) (13,33.3)
};
  \addlegendentry{FAIR}
  \addplot+[fill=gray!30, draw=black!70] coordinates {
    (0,0.0) (1,16.7) (2,0.0) (3,16.7) (5,0.0) (6,0.0) (7,15.8) (8,11.8) (10,11.1) (11,27.3) (12,54.5) (13,33.3)};
  \addlegendentry{WEAK}

\nextgroupplot[
  title={(b) Rankings of concerns raised by the \methodName obligations but not addressed by the baseline.},
  title style={font=\footnotesize, anchor=south west, at={(0,0.95)}, yshift=-0.5ex},
  ylabel={\%},
]  \addplot+[fill=blue!65!black, draw=black!70] coordinates {
(0,33.3) (1,28.0) (2,28.3) (3,17.6) (5,25.0) (6,25.0) (7,20.0) (8,27.8) (10,17.6) (11,21.7) (12,23.3) (13,23.8)
};
  \addplot+[fill=blue!30, draw=black!70] coordinates {
    (0,66.7) (1,60.0) (2,47.8) (3,58.8) (5,75.0) (6,43.8) (7,40.0) (8,55.6) (10,64.7) (11,65.2) (12,50.0) (13,76.2)
};
  \addplot+[fill=gray!30, draw=black!70] coordinates {
    (0,0.0) (1,12.0) (2,23.9) (3,23.5) (5,0.0) (6,31.2) (7,40.0) (8,16.7) (10,17.6) (11,13.0) (12,26.7) (13,0.0)};
\end{groupplot}

\foreach \x/\name in {0.153/HumanEval, 0.5/LiveCodeBench, 0.847/MBPP} {
  \node[anchor=north, font=\bfseries\small]
    at ($(cov c1r2.south west)!\x!(cov c1r2.south east)+(0,-1.5cm)$) {\name};
}
\end{tikzpicture}
\caption{\bf Usefulness of concerns identified exclusively by each approach. The baseline yielded 131 such concerns, compared to 343 for \methodName{}.}\label{fig:concerns}
\vspace{-4mm}
\end{figure}
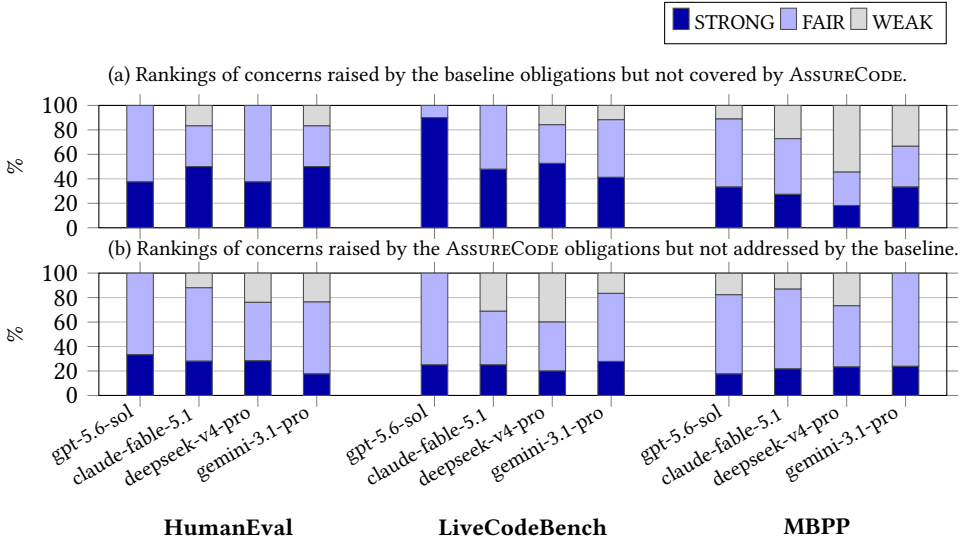
Fig.~\ref{fig:concerns} shows the concerns raised by only one of the two approaches, classified as ``Strong'', ``Fair'', or ``Weak'' by the judge. Across all configurations, \methodName{} raises 343 concerns not covered by the baseline (2.9 per task), compared with 131 baseline-only concerns (1.1 per task), and raises more unique concerns in 11 of the 12 configurations. The quality distribution is less asymmetric: baseline-only concerns are more frequently rated ``Strong'' (43\% vs. 24\%), while \methodName{}-only concerns are more concentrated in ``Fair'' (59\% vs. 41\%). Nevertheless, the proportion of concerns with at least potential assurance value is similar across the two approaches (84\% for baseline-only concerns and 83\% for \methodName{}-only concerns), and the proportion rated ``Weak'' is comparable (16\% vs. 17\%).
Because \methodName{} identifies more unique concerns overall, it still contributes more ``Strong'' concerns in absolute terms.
Note that we retain these results despite the bias discussed in Section \ref{sec:baseline-v-method}, as its impact is not large enough to affect our conclusions.
\begin{tcolorbox}[colback=gray!5,
  colframe=gray!40,
  boxrule=0.4pt,
  before skip=4pt,
  after skip=4pt,
  left=5pt,
  right=5pt,
  top=3pt,
  bottom=3pt]
\textbf{Answer to RQ4.}
Both approaches identify unique concerns with potential assurance value. \methodName{} raises substantially more unique concerns than the baseline, and approximately 83\% of these concerns are judged at least ``Fair''. Although baseline-only concerns are more frequently rated ``Strong'', \methodName{} still identifies more Strong concerns in absolute terms.
\end{tcolorbox}
\subsubsection{Summary of Observations }

Our goal was to compare the degree of assurance supported by \methodName{} with that supported by a vibe-justification baseline, and our results allow us to make three core observations. First, the two approaches largely converge on the concerns that should be considered when assessing the adequacy of generated programs. A single \methodName{} execution recovers most of the concerns surfaced by six merged baseline prompts, suggesting that the structured workflow does not substantially reduce the scope of assurance considerations. Second, the approaches are not interchangeable, as each identifies concerns that the other does not. However the coverage relationship is asymmetric: \methodName{} addresses a larger fraction of baseline concerns than the baseline addresses of \methodName{} concerns. This difference is partly explained by the formalization artifacts generated by \methodName{}, which resolve certain ambiguities that remain explicit concerns under unstructured prompting; however, formalization alone does not account for all of the observed coverage. Third, a larger number of obligations does not necessarily imply greater assurance value. The quantity and quality of unique concerns vary across models and benchmarks, and baseline-only concerns are somewhat more likely to receive the highest assurance-value rating. Thus, the contribution of \methodName{} is not uniformly producing more valuable obligations, but providing comparable assurance support through a more structured and systematic process. Together, these observations suggest that \methodName{} provides a viable structured alternative to vibe justification: it achieves comparable assurance support while replacing an ad hoc process of iterative prompting with an explicit methodology for generating and organizing assurance artifacts.

\paragraph{Threats to Validity}
Obligation count is only a proxy for scope: more obligations can mean more issues found, but also more review burden. Coverage also depends on obligation granularity, since a broad obligation is more likely to be judged as ``fully'' covering a concern than a narrow one, so coverage percentages partly reflect how each approach phrases its obligations. Usefulness (Strong/Fair/Weak) is an ordinal judgment, not evidence that a concern exposes a real defect. Finally, we do not measure effort or subjectivity directly; we simply argue that producing roughly equivalent sets of obligations via \methodName is a preferable alternative to ``vibe justification'' due to its more structured nature. The greatest threat is likely our use of an LLM judge. The judge, {\tt gpt-5.6-sol}, is also one of the four subject models, so it may favour obligations phrased in its own style. To mitigate known judgment and position biases~\cite{wang-etal-2024-large-language-models-fair}, the LLM judge sees the two obligation sets under anonymized labels and in random order, and we manually validate its output. These measures reduce but do not eliminate the risk, as phrasing may still reveal which set is which, and manual validation is prone to error and bias. Nondeterminism in \methodName executions could shift the results, as we only executed the tool once for each task. Our benchmarks (HumanEval, LiveCodeBench, MBPP) are small, self-contained, well-specified tasks, and results may not generalize to larger, more complex software projects.

\section{Discussion}
\label{sec:discussion}
\methodName{}, provides a structured mechanism for eliciting collateral validation obligations during AI-assisted software development. Rather than relying on an engineer to identify concerns through ad hoc inspection of generated artifacts, \methodName{} organizes these concerns within a broader argument structure that relates specifications, formalizations, verification artifacts, and residual questions requiring human judgment. Our evaluation provides initial evidence that this approach can achieve comparable assurance support to a practical vibe-justification baseline: a single \methodName{} execution recovered most concerns identified through multiple unstructured prompts while also surfacing additional concerns. Moreover, the coverage analysis highlights that this support arises from the complementary roles of formalization artifacts and validation obligations.

Several important aspects of properly evaluating \methodName{} remain open. Our evaluation considered a relatively small number of samples from three benchmarks consisting primarily of small, well-specified programming tasks; larger and more diverse datasets are needed to understand how the approach generalizes to more realistic software development settings. We also evaluated only four contemporary high-capability LLMs, and further investigation is needed to understand how the methodology behaves with smaller or less capable models. Finally, our evaluation prioritized the assurance artifacts and obligations produced by \methodName{} over the practical costs of producing them. Studies measuring token usage, developer effort, and the effect of \methodName{} on human decision-making and review practices would provide important evidence about its practicality. 

\section{Related Work}
\label{sec:related}

\paragraph{Trustworthy Agentic Software Development}
As AI-supported software development becomes increasingly commonplace, trust in AI-generated artifacts has received growing attention in software engineering. Roychoudhury et al.~\cite{roychoudhury2025agentic} identify nine \emph{trustworthiness factors} spanning human and technical concerns. Our current implementation of \methodName{} primarily addresses \emph{correctness} and \emph{explainability and trustworthiness}, and incorporates several of their identified contributors to trust: testing and static analysis, including validation of test oracles; formal proofs; and human--AI collaboration. Analysis and proofs are central to our design, as is human--AI collaboration, although the effectiveness of the latter has yet to be evaluated. Other contributors identified by Roychoudhury et al., including security guardrails and specification inference from software artifacts, are outside the current scope of \methodName{}, but could be incorporated within its general architecture. Khati et al.~\cite{khati2025mapping} similarly identify correctness and understandability as the concepts most frequently associated with trust in LLM-based software engineering. Brun et al.~\cite{brun2026automatically} propose a research roadmap combining interactive specification synthesis, agentic code generation, and static analysis and testing for trustworthy agentic development. At a high level, this corresponds to the activities integrated by \methodName{}; our contribution is to relate these activities through collateral validation obligations and organize their outputs into a structured justification rationale. \methodName{} was developed following an earlier proposal by some of the present authors in adapting assurance cases as explanations for the adequacy of AI-generated artifacts~\cite{murphy2026abductive}. The most significant difference between this proposal and \methodName{} is using the development of the justification rationale to \emph{drive} the generation of code, rather than attempting to argue for its adequacy \emph{post hoc}.
 
Existing frameworks and methodologies for agentic engineering use various strategies to support these aspects of trustworthiness, including constraining agent actions and providing execution feedback~\cite{yang2024swe}, and prescribing engineering workflows for agents to follow~\cite{xia2024agentless,hong2024metagpt,bouzenia2025repairagent}. TiCoder~\cite{fakhoury2024llm} is particularly close to \methodName{} in combining automated development with human validation.TiCoder generates candidate programs and tests, with users confirming or correcting individual tests to prune and rank candidates.   TiCoder also formalizes intent through instance-level tests and uses human feedback to select among candidate programs. In contrast, \methodName{} formalizes intent as a machine-checkable specification, uses it to verify the alignment of generated tests, and surfaces evidence adequacy and residual assumptions as explicit collateral validation obligations. Although our current prototype supports only functional testing, the approach is not limited in principle to testing as a source of evidence. To the best of our knowledge, \methodName{} is the first agentic development methodology to explicitly structure justification through argumentation inspired by assurance cases in safety-critical systems. Assurance cases have been widely adopted in safety-critical engineering, supported by a substantial literature on their design, analysis, and maintenance; \methodName{} is a first step towards adapting these techniques for agentic software engineering.

\paragraph{Formal Methods for Agentic Coding}
Formal methods have been identified as a promising component of trustworthy agentic software development~\cite{roychoudhury2025agentic,brun2026automatically}, as they can reduce ambiguities in natural-language prompts and provide machine-checkable guarantees that generated artifacts satisfy specified properties. Existing work explores several ways of incorporating these capabilities into LLM-based development. Endres et al.~\cite{endres2024can} study the ability of LLMs to translate natural-language tasks into formal postconditions that can be checked as program assertions. VeCoGen~\cite{sevenhuijsen2025vecogen} combines LLM-based program generation with compiler and formal-verification feedback to iteratively refine C programs against a provided specification. Other approaches integrate LLM-based code generation with proof assistants to verify correctness properties~\cite{mukherjee2025llm,tu2026agentic}. {FORGE}~\cite{wei2026formal} supports vibe coding in high-assurance domains through a closed-loop generation and verification process: LLM-generated Java is treated as an initial draft, transformed into several formal artifacts, and refined in response to verification failures translated into structured correction prompts. Like \methodName{}, FORGE combines generation, verification, structured feedback, and backtracking, but the approaches differ in how they handle the assumptions underlying this process. FORGE derives its formal artifacts from the generated implementation, whereas \methodName{} formalizes intent from natural language, and makes the fidelity of formalization a validation obligation.

\paragraph{Agentic Assurance Case Development}
 Odu et al.~\cite{odu2025automatic} automate assurance-case creation by instantiating structured assurance-case patterns, although these patterns are primarily \emph{structural} rather than the semantically grounded templates employed by \methodName{}. Trusta~\cite{chen2025trusta} uses LLMs to construct and verify Trustworthiness Derivation Trees (TDTs), an assurance-case-like formalism that, like our rationales, supports formal/informal expression pairs and automated reasoning over argument structures. Viger et al.~\cite{viger2026evaluating} investigate LLM-generated \emph{defeaters} for scrutinizing and refining assurance cases. These works primarily seek to improve safety-assurance practice using LLMs, whereas \methodName{} adapts assurance-case techniques to agentic software development. These efforts are complementary: insights from established assurance practice can inform agentic development (as in this work), while agentic development provides a new setting in which assurance case techniques can be extended and evaluated.
\section{Conclusion}
\label{sec:conclusion}
In this paper, we introduced justification rationales for agentic coding that make explicit the claims made about generated code, the evidence offered for them, and the collateral validation obligations a developer should address before trusting that code. We described \methodName, a tool-supported  methodology that creates such rationales during agentic code generation through autoformalization, theorem proving, and agentic artifact generation. Our evaluation suggests that \methodName{} provides comparable assurance support to prompt-based \emph{vibe justification}, while using partial formalization and structured argumentation to systematically generate and organize validation obligations. These results support \methodName{} as a credible option for building correctness assurance in AI-assisted software development.

Several directions for future research follow from this work. First, user studies are needed to assess the effectiveness of \methodName in practice, including how it affects developers' understanding of and trust in generated code, and to identify practical barriers to adoption. Second, we plan to extend justification rationales beyond functional correctness to design constraints, nonfunctional and normative requirements, and security. Third, we plan to apply \methodName to larger software projects, especially where assurance concerns interaction between components.

\bibliographystyle{acm}
\bibliography{refs}

\end{document}